\documentclass[apl,noshowpacs,preprint]{revtex4}
\usepackage{mathrsfs}
\usepackage{graphicx}
\usepackage{dcolumn}
\usepackage{bm}
\usepackage{txfonts}

\begin{document}

\title{Dependence of transport on adatom location for armchair-edge
graphene nanoribbons}

\author{Xiongwen Chen,$^{1,2}$ Kehui Song,$^2$ Benhu Zhou,$^1$
Haiyan Wang$^1$}
\author{Guanghui Zhou$^{1,3}$}
\email{ghzhou@hunnu.edu.cn}

\affiliation{$^1$Department of Physics and Key Laboratory for
Low-Dimensional Quantum Structures and Manipulation (Ministry of
Education), Hunan Normal University, Changsha 410081, China}

\affiliation{$^2$Department of Physics and Electronic Information
Science, Huaihua University, Huaihua 418008, China}

\affiliation{$^3$International Center for Materials Physics, Chinese
Academy of Sciences, Shenyang 110015, China}

\begin{abstract}
We study the transport property for armchair-edge graphene
nanoribbons (AGNRs) with an adatom coupling to a semi-infinite
quantum wire. Using the nonequilibrium Green's function approach
with tight-binding approximation, we demonstrate that the tunneling
current through the system is sensitively dependent on both the AGNR
width and adatom location. Interestingly, when the adatom locates
onto a carbon atom in the 3$j$th chain from the edge of a metallic
AGNR, the system shows a transmission gap accompanied by a threshold
voltage in $I-V$ curve like a semiconducting AGNR. This effect may
be useful in scanning tunneling microscopy experimental
characterization on graphene samples.
\end{abstract}

 \maketitle

Recently, the successful fabrication of graphene in experiment$^1$
results in intensive attention due to its unique properties$^2$ and
potential applications in electromechanical resonator,$^3$ $p$-$n$
junction,$^4$ transistor$^5$ and thermoelectric power devices,$^6$
etc. However, due to the absence of gap in the energy spectrum,
graphene is difficult to be used in the electronic device which
requires precise control of carrier type and transport behavior.
Fortunately, GNRs, finite width of graphene, open a gap in the
energy spectrum because of the quantum confinement effect,$^{2,7}$
remedying the drawback of graphene. Thus, GNRs are more suitable
promising materials for the design of electronic devices.$^{8-12}$
According to the atomic configuration of the edge, GNRs are
classified into two types of AGNR and zigzag-edge GNR (ZGNR). It has
been shown that a pristine AGNR exhibits either metallic or
semiconducting nature with a gap depending on the number of
elementary cells $N$ from one edge to another,$^{13,14}$ i.e., an
AGNR is metallic if $N$=3$p$-1 ($p$ is an integer) otherwise
($N$=3$p$-2 or 3$p$-3) semiconducting, while ZGNRs always maintain
metallic behavior with localized states near the Fermi level.$^{13}$

In this Letter, we study the transport property for AGNRs with an
impurity atom absorbed on the top of a carbon atom in AGNR. As shown
in Fig. 1(a), like the model proposed for STM experiment on
graphene$^{15-18}$ or surface of materials,$^{19}$ a semi-infinite
quantum wire (as a probe) couples to an AGNR via an adatom. In Fig.
1(b), the (red) dashed line rectangle is the periodical unit which
is composed of $N$ elementary cells for AGNR,$^{13,14}$ where cell
$(m,n)$ contains four carbon atoms labeled as $\beta,\gamma,\lambda$
and $\delta$ connected by the (blue) thick solid line. Using the
NEGF approach with the help of tight-binding approximation, we
demonstrate that the tunneling current through the system is
sensitively dependent on both the width of AGNR and the location of
the adatom on it. Interestingly, when the adatom locates onto a
carbon atom in the 3$j$th chain from the edge of a metallic AGNR,
system shows a transmission gap accompanied by a threshold voltage
in the current-voltage curve, which is a conspicuous semiconducting
transport behavior in the semiconducting-AGNR-based system. This
effect may be useful in the STM experimental characterization on the
GNR samples and in the application of graphene-based nanodevices.

The Hamiltonian for this AGNR-based system reads
\begin{eqnarray}
H=H_{W}+H_{G}+H_{I}+H_{T},
\end{eqnarray}
where
$H_{W}$$=$$\sum_{j=1}^{\infty}\mu_{1}a^{\dag}_ja_j$$-$$\sum_{j=1}^{\infty}t_{1}(a^{\dag}_ja_{j+1}$$+$$h.c.)$
for the semi-infinite wire with chemical potential $\mu_{1}$ and
hopping energy $t_{1}$;
$H_{G}$$=$$\sum_{m,n,\alpha}\mu_{2}c^{\dag}_{m,n,\alpha}
c_{m,n,\alpha}$$-$$\sum_{<\cdot,\cdot>}t_2(c^{\dag}_{m,n,\alpha}c_{m',n',\alpha'}+h.c.)$
for the $\pi$-band electron of AGNR with chemical potential
$\mu_{2}$, where $<$$\cdot,\cdot$$>$ denotes summing over
nearest-neighbor sites with hopping integral $t_2$;
$H_{I}$$=$$\epsilon_0d^\dag d$ for the adatom with single energy
level $\epsilon_0$ couples to both site $j$=1 of the wire and
lattice site $\alpha$ in AGNR respectively through coupling strength
$\nu_1$ and $\nu_2$ with tunneling Hamiltonian
$H_{T}$$=$$(\nu_1d^\dag a_1+\nu_2d^\dag c_{m,n,\alpha}+h.c.)$. The
potential difference between the wire and AGNR (via substrate) is
related to the applied bias voltage as $U$=$\mu_1$$-$$\mu_2$.

Therefore, for the AGNR-based system of an adatom located in cell
$n$, the tunneling current can be expressed as
\begin{eqnarray}
I_n(U)=\frac{2e}{h}\int_{\mu_2}^{\mu_1}d\omega[f_1(\omega-\mu_1)-f_2(\omega-\mu_2)]T_n(\omega),
\end{eqnarray}
where $f_{1(2)}$ is the Fermi-Dirac distribution function for the
wire (AGNR), $T_n$ is the $n$-dependent transmission probability
which can be obtained by solving full Hamiltonian (1) by means of
NEGF within the nearest-neighbor tight-binding scheme.

The isolated retarded Green's function (GF) matrix for the system is
defined as
\begin{eqnarray}
g^{r}(\omega)={\begin{array}{*{20}c}\end{array}}\left[
{\begin{array}{*{20}c}g^{r}_{11}(\omega)&&&0&&&0\\0
&&&g^{r}_{00}(\omega)&&&0\\0&&&0&&&g^{r}_{m,n,\alpha}(\omega)
\end{array}}\right],
\end{eqnarray}
where $g^{r}_{00}(\omega)$$=$$(\omega-\epsilon_0+i0^{\dag})^{-1}$ is
the GF for the adatom, $g^{r}_{11}(\omega)$ and
$g^{r}_{m,n,\alpha}(\omega)$ are GF for the edge site ($j$=1) of the
wire and site $\alpha$ in cell $n$, respectively. Using the NEGF,
one finds$^{16,17}$
\begin{eqnarray}
g^r_{11}(\omega)=\frac{\omega-\mu_1}{2t^2_1}-i\frac{\sqrt{4t^2_1-(\omega-\mu_1)^2}}{2t^2_1}
\end{eqnarray}
with local density of sates (LDOS)
$\rho_{11}(\omega)$$=-\frac{1}{\pi}Im [g^r_{11}(\omega)]$ at site
$j$=1 in the wire and
\begin{eqnarray}
g^r_{m,n,\alpha}(\omega)=\sum_{k_x,k_q,\pm}\frac{\psi_{m,n,\alpha}(k_x,
k_q)\psi_{m,n,\alpha}^*(k_x,k_q)}{\omega-E^{\pm}+i0^\dag}
\end{eqnarray}
with LDOS $\rho_{m,n,\alpha}(\omega)$$=$$-\frac{1}{\pi}Im[
g^r_{m,n,\alpha}(\omega)]$ at site $\alpha$ in the $n$th cell of
AGNR.

Further, based on the nearest-neighbor tight-binding approximation,
one can obtain$^{13,14}$ the $\pi$-electron energy spectrum in the
low-energy limit as
\begin{eqnarray}
E^{\pm}=\mu_2\pm
t_2\sqrt{1-4\text{cos}\frac{k_q}{2}\text{cos}\frac{3k_x}{2}+4\text{cos}^2\frac{k_q}{2}},
\end{eqnarray}
and the wavefunction
\begin{eqnarray}
\psi_{m,n,\alpha}(k_x,k_q)=C\left\{\begin{array}{l
l}e^{i\varphi_\alpha(k_x,k_q)}\text{sin}(k_qn),\,\,\,\,\,\,\,\,\,\,\,\,\,\,\,\,\,\,\,\alpha=\beta,\gamma\\
e^{i\varphi_\alpha(k_x,k_q)}\text{sin}[k_q(n-1/2)],\,\alpha=\lambda,\delta
\end{array} \right.
\end{eqnarray}
under the hard-wall boundary condition with normalized constant
$C$=$[2(N+1)]^{-1/2}$, where $\varphi_{\alpha}$ is an arbitrary
function and the carbon-carbon bond length has been set to unitary.
The wavevector $k_x$ is continuous and the $y$-direction wavevector
is discretized as $k_q$=$q\pi/(N+1)$ with subband index number
$q$$=$$1,2,\cdots,N$ for a perfect AGNR. Hence, the energy gap
$\Delta E$ between the $q$th conduction and valence band is given by
$2t_2|1-2\text{cos}(k_q/2)|$.

Moreover, the retarded self-energy matrix has four nonzero elements
of $\Sigma_{1,0(0,1)}^r$=$\nu_1$
($\Sigma_{0,\alpha(\alpha,0)}^r$=$\nu_2$), which describes the
tunneling between the adatom and wire (AGNR), respectively.
Therefore, by employing Dyson equation
$G^{r}$=$[(g^{r})^{-1}$-$\Sigma^{r}]^{-1}$, one gets the full
retarded GF for the adatom $G^r_{00}(\omega)$ and lastly obtains the
electronic transmission probability
$T_n$=$4\Gamma_{01}(\omega)|G^r_{00}(\omega)|^2\Gamma_{02}(\omega)$
with linewidth functions $\Gamma_{01}(\omega)$=$\pi
|\nu_1|^2\rho_{11}(\omega)$ and $\Gamma_{02}(\omega)$=$\pi
|\nu_2|^2\rho_{m,n,\alpha}(\omega)$.

In the following we present some numerical examples for the systems
with $N$=7 and 8 AGNR, respectively. The potentials
$\mu_1$=-$\mu_2$=$U/2$ and the hopping energies $t_1$$=$$t_2$=$t$,
where we have set the origin of energy as the Fermi energy ($E_F$=0)
at equilibrium case. In our model, when the adatom energy
$\epsilon_0$ deviates from $E_F$ along the positive or negative
direction, the impurity scattering for electronic transmission
suppresses the amplitude of tunneling current. But this do not
change the major characteristics for electronic transmission which
is determined mainly by the energy spectrum of AGNRs. Thus, to
obtain the essence of transport property for our system, we fix
$\epsilon_0$ at $E_F$ and set the coupling strength
$\nu_1$$=$$\nu_2$=0.5 without loss generality.

In Fig. 2, transmission probability $T_n$ versus energy $\omega$ (in
units of $t$) at equilibrium case of $\mu_{1(2)}$=0 and tunneling
current $I_n$ (in units of $e/h$) versus bias voltage $U$ for
different adatom positions are depicted, respectively. Obviously,
the transport property of the system strongly depend on the width
(or number of cells $N$) of AGNR and the adatom location ($n$) on
it. As shown in Fig. 2(a) for the system with $N$=7, the
semiconducting nature is always kept regardless of $n$, which is
signed by a transmission gap $\Delta\omega$$\approx$ 0.22 around
$E_F$=0 and a series of signature peaks from subbands. This gap can
be determined from the minimal energy gap $\Delta E_{min}$.
Consequently, as shown in Fig. 2(b), only when the applied bias
voltage $U$ is above $\Delta\omega/2$, carriers in the highest
valence band ($q$=5) can jump to Fermi level and take part in
transport, thus there exists a threshold voltage $U_c$$\approx$0.11
for the semiconducting system. As $U$ increases, other subbands
become propagating accordingly one by one, which results in
broken-line signs in $I$-$U$ curve in contrast to the smooth $I$-$U$
characteristic in the previously studied similar graphene-based
system.$^{16,17}$

In contrast, for $N$=8 the system shows a metallic behavior with a
zero transmission gap [see Fig. 2(c)] and linear $I$-$U$
characteristics [Fig. 2(d)] when the adatom is located at site
$\beta$ or $\gamma$ in cells of $n$=1,2,4 because that the sixth
($q$=6) valence band and conduction band meet at the Fermi energy
with a zero gap.$^{11}$ Interestingly, when the adatom is located at
cell $n$=3, there exists a wide transmission gap of
$\Delta\omega$$\approx$0.54 associated with a large threshold
voltage $U_c$$\approx$0.27 for the system [see (black) solid lines
in Fig. 2(c) and 2(d)]. In this case, the metallic-AGNR-based system
surprisingly exhibits semiconducting behavior. Actually, this
phenomenon can be explained by wavefunction (7). As shown by the
(black) solid line in Fig. 3(a), for metallic AGNR with $N$=8, the
adatom locating at site $\beta$ or $\gamma$ in cell $n$=3 results in
$|\psi_{m,3,\beta(\gamma)}(k_x, k_6)|^2$=0, hence the sixth
(gapless) subband does not contribute to transport in this case.
Generally, for any metallic-AGNR-based system, the semiconducting
transport behavior exists only if the adatom is located at the
positions of carbons in cells of $n=3j$ for $\alpha=\beta$($\gamma$)
and $3j-1$ for $\alpha=\lambda$($\delta$) with integer $j$,
respectively. In other words, the zero gap subband is not effective
when the adatom is located at the 3$j$th chain from the edge of AGNR
because $|\psi_{m,n,\alpha}|^2$=0 as $q=2(N+1)/3$. Comparatively,
$|\psi_{m,n,\alpha}|^2$ at the $q'$th subband with $q'$=$[2(N-1)/3]$
(where $[\cdots]$ rounds a number to the nearest integer) begins to
have nonzero value from the Fermi level. Therefore, the threshold
voltage $U_c$ is taken as $t_2|1-2\text{cos}[q'\pi/(2N+2)]|$
depending on the metallic AGNR width $N$. Of course, this transition
from metallic to semiconducting does not exist for either metallic
AGNRs transport in the graphene-plane$^{9-12}$ or graphene-based
systems.$^{15-18}$

On the other hand, as an application of this transition effect found
here, firstly one can find the location of an adatom on an AGNR
sample in a STM experiment. Further, for a set of AGNRs with same
$p$, the metallic AGNR possesses the widest $\Delta E_{min}$ when
its zero gap subband is not absent as shown by the (black) solid
line in Fig. 3(b). Therefore, when $\Delta E_{min}$ is small enough
to be neglected, all AGNRs can be treated as graphene. The turning
point is around $N$=400 which corresponds the width of AGNR about
$100$ nm. Moreover, because of quantum confinement, the difference
between energy levels in the wider AGNRs is smaller than that of
narrower AGNRs. So in the wider AGNRs, the number of conduction
channels within bias voltage window changes less abruptly, which
gives rise to a smooth $I$-$U$ curve as shown with the (black) solid
lines in Fig. 4(b) and 4(d) despite the existence of weak
oscillating in transmission spectrum displayed in Fig. 4(a) and
4(c). Therefore, the smooth $I$-$U$ curve is an visible symbol for
the wider AGNRs compared to the broken-line signs in $I$-$U$ curve
for the narrower AGNRs [see (blue) dot and (red) dashed lines in
Fig. 4(b) and 4(d)]. However, a threshold voltage $U_c$ (decreases
as $N$ increases) always exists when $n$=$3j$ for
$\alpha$=$\beta$($\gamma$) and $3j-1$ for
$\alpha$=$\lambda$($\delta$), respectively, even though $N$ is large
enough to neglect the finite-size effect [see Fig. 4(d)] in our
model. Therefore, one can use the magnitude of $U_c$ to identify an
AGNR sample in practical STM experiment by our setup shown in Fig.1.

In conclusion, we have considered an AGNR-based system with a single
adatom couples to both a semi-infinite wire and an AGNR. Using the
NEGF technique within the nearest-neighbor tight-binding scheme, it
is found that the tunneling current of the system is sensitively
dependent on both the width of AGNR and the location of adatom on
it. Interestingly, when the adatom is on a carbon atom in the 3$j$th
chain from the edge of a metallic AGNR, a metallic-AGNR-based system
always exists a transmission gap and consequently followed by a
threshold voltage $U_c$ in $I$-$U$ curve, which is a conspicuous
semiconducting transport behavior in the semiconducting-AGNR-based
system. Although the values of transmission gap and $U_c$ in real
situation may slightly differ from our theoretical prediction due to
the unsaturated edge bonds or other factors which induce the
variation of subbands for GNRs,$^{8-12}$ we believe that the main
point here is qualitatively sound. And this effect may be useful in
the STM experimental characterization on AGNRs and in the
application of graphene-based nanodevices.

We thank Dr. Dongsheng Tang for insightful discussions. This work
was supported by the National Natural Science Foundation of China
(Grant No. 10974052), the Scientific Research Fund of Hunan
Provincial Education Department (Grant No. 09B079), and the Program
for Changjiang Scholars and Innovative Research Team in University
(PCSIRT, No. IRT0964).

\newpage
{\bf List of Figure Captions:}
\baselineskip=27pt

Fig.1: (Color online) (a) Electrons tunneling from a semi-infinite
wire to an AGNR on substrate via an (red) adatom on a carnon site of
AGNR. (b) The (red) dashed line rectangle represents the $m$th unit
along the $x$-direction, in which the elementary cell $n$ along the
$y$-direction is composed of four carbon atoms
$(\beta,\gamma,\lambda,\delta)$ connected by blue line.

Fig.2: (Color online) Transport property of the system with
different AGNR width $N$ and the adatom position $n(\beta,\gamma)$.
Left for transmission probability $T_n(\omega)$ and right for
tunneling current $I_n(U)$, where (a) and (b) for $N$=7, (c) and (d)
for $N$=8.

Fig.3: (Color online) (a) Squared wavefunctions for different
subband indices in metallic AGNR of $N$=8 as a function of adatom
position. (b) Minimal energy gap as a function of AGNR width. Note
that the zero energy gap in pristine metallic AGNR ($N$=$3p-1$) is
absent in the corresponding adatom-attached ribbon as indicated by
(black) solid line.

Fig.4: (Color online) Transport property for metallic-AGNR-based
system with different $N$ and $n$, where (a) and (b) for $n$=1, (c)
and (d) for $n$=3.

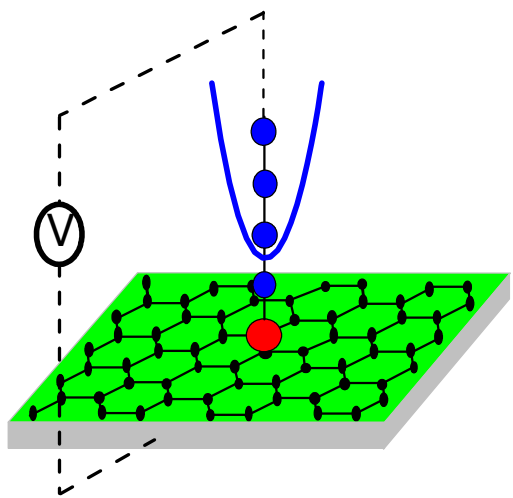
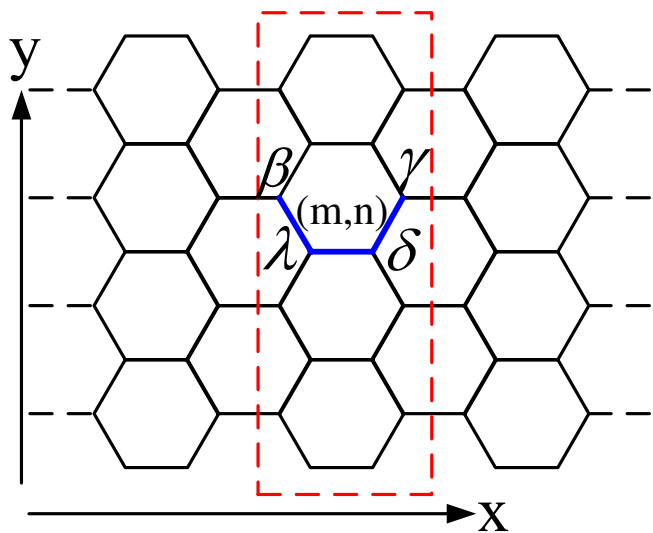
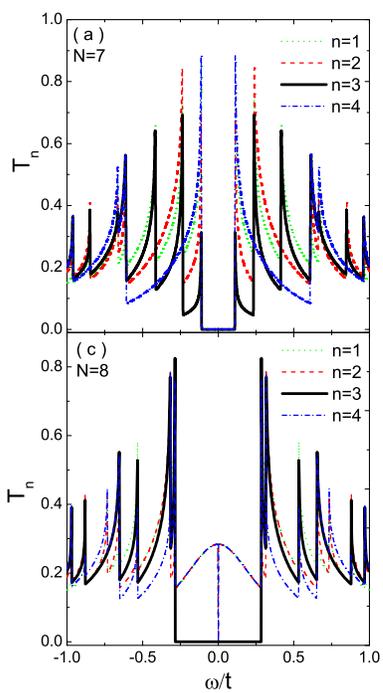
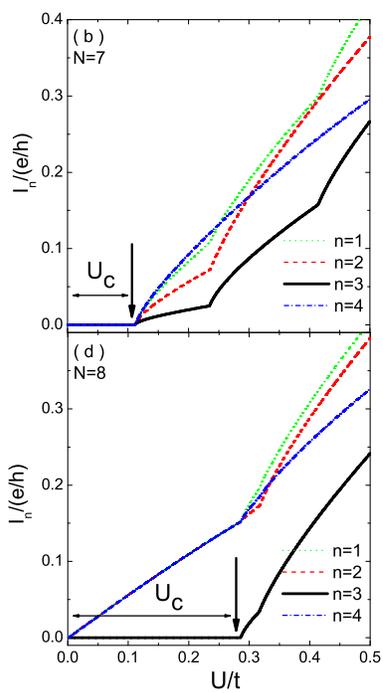
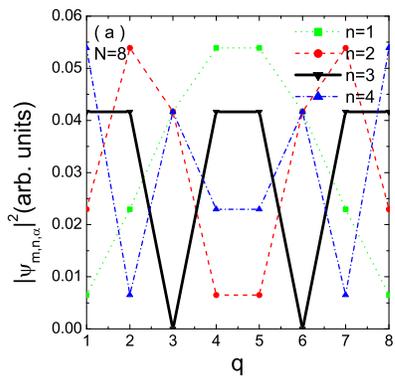
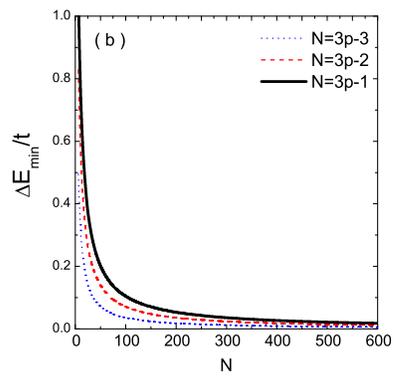
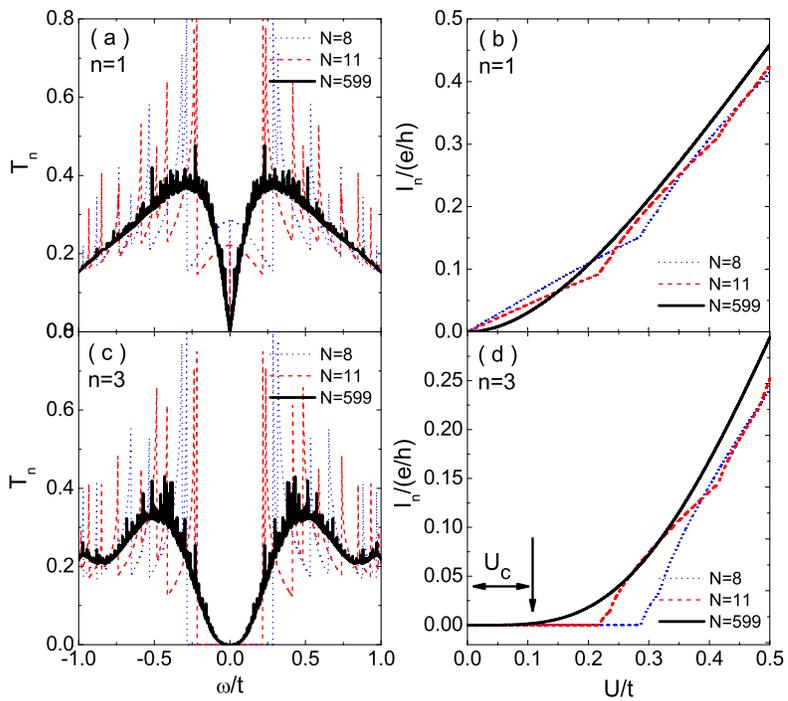
\end{document}